\documentstyle[aps,multicol,prl]{revtex}

\tighten

\begin{document}

\pagestyle{empty}

%\begin{flushright} hep-th/0204160\\UCB-PTH-02/18\\LBNL-51485\\CMU-HEP-
%\end{flushright}

%\end{minipage}

\title{High Energy Field Theory in Truncated AdS Backgrounds}
%HighEnergy Behavior of Gauge Fields in Compactified AdS Models}
\author{Walter D. Goldberger$^{1,2}$\thanks{walter@thsrv.lbl.gov} and
Ira Z. Rothstein$^3$\thanks{ira@cmuhep2.phys.cmu.edu}}
\address{$^1$Department of Physics, University of California,
Berkeley, CA 94720}
\address{$^2$Theoretical Physics Group, Lawrence Berkeley National
Laboratory, Berkeley, CA 94720}
\address{$^3$Department of Physics, Carnegie Mellon University, 
Pittsburgh, PA 15213}\maketitle

\begin{abstract}
In this letter we show that
in five-dimensional anti-deSitter space (AdS) truncated by boundary
branes, effective field theory techniques are reliable at high energy
(much higher than the scale suggested by the Kaluza-Klein mass gap),
provided one computes suitable observables.   We argue that
in the model of Randall and Sundrum for generating the weak scale
from the AdS warp factor, the high energy behavior of gauge fields
can be calculated in a {\em cutoff independent manner}, provided one
restricts Green's functions to external points on the Planck brane.
Using the AdS/CFT correspondence, we calculate the one-loop
correction to  the Planck brane gauge propagator due to
charged bulk fields.   These effects give rise to non-universal logarithmic
energy dependence for a range of scales above the Kaluza-Klein gap.

\end{abstract}

\begin{multicols}{2}[]

\pagestyle{plain}
\narrowtext

\section{Introduction}

It is widely believed that there exists a large energy desert
separating the weak scale from the scale of grand unification, or
perhaps the Planck scale.  A primary reason for this belief is the
observation that high-scale perturbative unification seems plausible
in supersymmetric extensions of the Standard Model.  On the other
hand, it appears that in models in which the weak scale is
fundamental, such as scenarios based on extra dimensions, one
has to abandon the idea of gauge unification at high energies.

Several authors have argued that
 in scenarios where the weak/Planck 
hierarchy is generated by a single warped 
extra dimension~\cite{RS1} (the RS model), it is possible for bulk gauge 
couplings to evolve logarithmically over a large range of scales~\cite{rsrun}~\cite{pomarol,choi}.  This is in sharp contrast to the behavior of a gauge theory 
propagating on compactified 5D flat space 
(for instance flat $R^4\times S^1$), where
 vacuum polarization effects give rise to power law 
corrections that become dominant at mass scales 
comparable to the compactification scale.  At the scales where 
these power corrections dominate, 
effective field theory techniques break down.  
In order to make predictions beyond such energies, 
local field theory must be embedded into a suitable 
UV completion that resolves its short distance singularities.

While we agree that high energy logarithmic behavior 
is possible in gauge theories propagating 
in compactified AdS, we find that this is physically 
realized in a manner that is somewhat different from 
the proposal of~\cite{rsrun}.  We will show that as in flat space, 
the one-loop correction to the gauge field 
zero mode propagator in compactified AdS contains
 a power law term that is saturated at 
the Kaluza-Klein mass scale (of order TeV).  
This power correction is a finite, 
non-analytic function of the 4D momentum.  Because of this, 
it cannot be removed by any procedure for regulating the 
ultraviolet divergences that arise in loop calculations.  
However, while in a flat space theory the 
emergence of power law behavior at the KK mass scale 
signifies a breakdown of the higher dimensional field theory description, 
the same is not true for theories 
propagating in background AdS spaces.  
Rather, in AdS, large loop corrections 
at the TeV scale imply a breakdown of the field theoretic
description of the zero mode observable, but not necessarily of other
correlators.  Thus, high energy gauge theory effects may be accessible
via local
field theory, provided one calculates the appropriate observable. 
 
In the RS proposal for obtaining the weak scale 
from the AdS warp factor~\cite{RS1}, a set of such calculable 
quantities is the 5D gauge theory correlators 
with external points restricted to the Planck brane.  
On the Planck brane the local cutoff for correlators is of order 
the AdS curvature scale (which is taken to be of order the Planck
scale and is much larger than the KK scale).  
This implies that it is this scale, 
and not the TeV scale, which suppresses power corrections 
to these observables.  Thus the logarithms due to KK zero modes in
loops give the dominant non-analytic corrections for a large range of
scales. 

To illustrate these claims, in this letter we will work in the context of 5D massless scalar electrodynamics as a model of gauge dynamics.  In principle, 
calculations in the model can be done in 5D, 
in a manner that is independent of any specific regulator.  
However, in order to compute the correlators on the Planck brane, it is easier to employ the AdS/CFT correspondence~\cite{maldacena,AdSCFT} as it 
applies to models with boundary branes~\cite{APR,RZ,PV}.  
The Planck brane vacuum polarization of our 
toy model can be computed in a dual 4D field theory 
that consists of a massless scalar and a 
gauge field which are both weakly coupled 
to a (broken) conformal field theory (CFT).  
The one-loop logarithm of the 5D theory 
can be calculated in this framework without
any detailed knowledge of the CFT
\footnote{This one-loop logarithm should not 
be confused with the classical high energy 
logarithmic behavior of the gauge field 
propagator on the Planck brane.  
That tree-level logarithm (attributed to CFT 
effects in the dual 4D theory) is universal, 
and would not be observed in the relative 
evolution of different gauge couplings at high energy.}.  

At energies of order the KK mass gap, the 
Planck brane correlators match on to the zero mode Green's functions.
Therefore, there is a calculable relation 
between high energy couplings and the parameters 
measured in low energy experiments.  
In particular, symmetry constraints on the dynamics 
near the curvature scale (for instance, 
high energy unification of gauge forces) 
could have meaningful implications for low energy physics.

Our setup is as follows.  We will work in the context of field theory
propagating in a background five-dimensional (Euclidean) 
AdS spacetime with the coordinatization
\begin{equation}
\label{eq:metric}
ds^2 = G_{MN} dX^M dX^N={1\over (kz)^2}\left(\eta_{\mu\nu} dx^\mu
dx^\nu + dz^2\right),
\end{equation}
where $z$ parameterizes location in the AdS bulk.  The AdS boundary
is cut off by a Planck brane at $z=1/k$, and the AdS horizon removed by the
presence of a TeV brane at $z=1/T.$  Here $k$ is the AdS curvature
parameter, and $T$ is an energy scale that sets the masses of KK
excitations of bulk fields.  For applications to the
hierarchy problem all parameters are taken to scale as appropriate
powers of the Planck scale, except $T\sim\mbox{TeV}.$  

\section{Flat Space Effective Field Theory and Unification}

Let us now review the one-loop structure of gauge theory compactified
on spaces of the form $R^4\times S^1$ or $R^4\times
S^1/Z_2$~\cite{flat}.  We will work in the context of massless scalar electrodynamics in five-dimensions.  This simple model retains the essential physical
features that we wish to discuss while avoiding technicalities that
arise in more realistic models of non-Abelian gauge fields and
fermions.  It is convenient to work with an observable that matches onto measurable 4D quantities at low energies (smaller than the compactification radius) .  Such an observable is provided by the two-point
correlator of the zero mode of the 5D gauge field, which is generated
by a 5D effective action evaluated on gauge fields that have no
dependence on the compact coordinate.  Keeping only terms up to
quadratic in the gauge field and working in the $A_5=0$ gauge
 this is given by
\begin{equation}
S_{eff} = {1\over 2}\int {d^4 q\over (2\pi)^4} A_\mu(q) \left[q^\mu q^\nu - q^2 \eta^{\mu\nu}\right]\Pi(q^2) A_\nu(-q),
\end{equation}
On $R^4\times S^1$, summing the tower of scalar KK modes running in the loops, and using dimensional regularization to regulate the short distance divergences yields 
\begin{eqnarray}
\label{eq:S1}
\nonumber
\Pi_{S^1}(q^2) &=& {1\over g_4^2} - {R\sqrt{q^2}\over 256}\\
& & {}  - {1\over 8\pi^2} \int_0^1 dx x\sqrt{1-x^2} \ln\left[1-e^{-\pi x R\sqrt{q^2}}\right].
\end{eqnarray}
The first term in this equation is simply the tree level action, with
$g_4$ the coupling of the zero mode gauge field to the tower of KK
states.  On the circle, this is related to the 5D gauge coupling by
$1/g_4^2 = 2\pi R/g_5^2$.  The second two terms represent the one-loop
contributions.  We have written the one-loop corrections as a piece
which is identical to the vacuum polarization of an uncompactified 5D
gauge theory (with zero external momentum in the fifth direction), and
a piece which contains finite radius effects.

Note that on $R^4\times S^1$, the vacuum polarization is finite.
Since $[g_5]=-1/2,$ divergent contributions to the zero mode vacuum
polarization in a massless theory must scale as $g_5^2 \Lambda$, with
$\Lambda$ an ultraviolet cutoff. However, we have used dimensional
regularization to regulate divergences, which simply sets these pure
counterterm contributions to zero.  
Indeed, from the 5D point of view
the concept of running couplings is not particularly meaningful, since
the resummation of power corrections is made moot by the fact
that when these contributions are of order one calculability is lost. 

Although there are no ultraviolet logarithms, finite logarithms can
arise from the infrared region of loop integrals.  For $\sqrt{q^2}
R\ll 1,$ Eq.~(\ref{eq:S1}) becomes
\begin{equation}
\Pi_{S^1} (q^2) \simeq {1\over g_4^2} - {1\over 24\pi^2}
\left[\ln\left(2 \pi R\sqrt{q^2}\right) -{4\over 3}\right]. 
\end{equation}
Taking the limit $R\rightarrow 0$ with $g_4$ fixed, we see that this
finite logarithm becomes singular.  But this is precisely the 4D
limit, in which the massive KK sates decouple.  Thus the singular
$R\rightarrow 0$ limit manifests itself as the usual one-loop UV
logarithm of a 4D theory.  The large logarithms that arise in this
regime can be resummed by using the renormalization group in 4D.
Besides the logarithm that appears in this equation, the expansion for
$\sqrt{q^2} R\ll 1$ of the integral in Eq.~(\ref{eq:S1}) also yields a
non-analytic $\sqrt{q^2} R$ piece as well as a series of integer
powers of $q^2$.  The non-analytic power-law term generated by this
expansion cancels exactly the explicit power law that appears in
Eq.~(\ref{eq:S1}), leaving behind the logarithm plus an analytic
function of $q^2$.  This is exactly what we expect for the behavior of
the zero mode correlator if the low energy limit of the 5D
compactified theory is to be reproduced by an effective 
4D theory of massless modes plus a tower of non-renormalizable local
operators. 

For later comparison with the compactified AdS theory, we also consider the zero mode two-point correlator for flat field theory on a finite interval.  Taking the space to be $R^4\times S^1/Z_2$ and making the bulk scalar even under the orbifold action, one finds in this case:
\begin{eqnarray}
\label{eq:orb}
\nonumber
\Pi_{S^1/Z_2}(q^2) &=& 
 {1\over 48\pi^2}\left[{1\over\epsilon} +\ln(4\pi) -{\gamma\over 2}
+{4\over 3}\right]\\
& & {} - {1\over 96\pi^2}\ln\left({q^2\over\mu^2}\right)+ {1\over
2}\Pi_{S^1}(q^2),
\end{eqnarray}
where $\mu$ is a subtraction scale.  In contrast to the result for $S^1$, the orbifold calculation encounters logarithmic divergences, represented here by a $1/\epsilon$ pole.  This is because the presence of the orbifold fixed planes has modified the divergence structure of the field theory.  Indeed, it is
now possible to write down brane localized gauge kinetic operators 
\begin{equation}
{\cal L} = {1\over 4}\left[\lambda_0\delta(z)  + \lambda_1\delta(z-\pi
R)\right] F_{\mu\nu} F^{\mu\nu} 
\end{equation}
with $\lambda_{0,1}$ dimensionless.  By dimensional analysis, a
divergent one-loop correction to these boundary couplings depends on
an ultraviolet cutoff as  $\ln\Lambda$, which corresponds to the
$1/\epsilon$ pole in Eq.~(\ref{eq:orb}).  Thus loop effects induce
non-trivial RG flows for boundary couplings.

How do we reconcile the fact that 5D bulk couplings do not run with
the statement of ``power-law running'' and unification that is often
made in the literature~\cite{dds}?  In general, gauge unification implies
that above some energy scale $M_G$ associated with symmetry breaking
of a unified gauge group, the forces mediated by the exchange of
unbroken gauge bosons are equal in magnitude.  In our 5D gauge
theories, this equality should hold for the full gauge two-point
correlators evaluated at arbitrary external positions in the fifth
direction\footnote{However, in models where GUT symmetry is broken by boundary conditions~\cite{oguts}, the equality of correlators evaluated on surfaces of broken symmetry will be spoiled by corrections that are small on the basis of naive dimensional analysis.}.  In particular, averaging the 5D correlators over the compact direction generates a condition on the zero mode vacuum
polarization of the unbroken gauge bosons at the unification scale
(assuming, for instance, a symmetry breaking pattern $G\rightarrow
G_1\times G_2$)
\begin{equation}
\label{eq:GUT}
\Pi^{G_1}(M_G^2) = \Pi^{G_2}(M_G^2).
\end{equation}
This equation can be used to obtain a relation between the gauge
couplings in the 5D Lagrangian.  From this, one can extract
 predictions for low energy observables, leading to the usual
prediction for the couplings at the weak scale
with power law running. However, 
one might expect that unknown UV effects could be large and put into
question the reliability of predictions when the power law starts to
play an important role.

For both the 5D circle and orbifold theories, the non-analytic power
corrections arising from the ultraviolet parts of 5D loop integrals
dominate at a scale which is, up to a phase space factor, of order
$1/R.$  At such energies, the 5D flat space gauge theory becomes
strongly coupled and ceases to be a useful effective description of
the physics.  While in flat space all Green's functions associated
with the effective gauge theory exhibit this type of behavior, the
same is not necessarily true in curved backgrounds, such as AdS.
Then, the relevant question for compactified AdS is, for a given
observable, at what scale does power law behavior become the dominant
effect?

\section{Effective Field Theory  and Observables In AdS}
We would like to understand which AdS 
observables can be studied in the framework of 
effective field theory (EFT) at a given energy scale.  
Consider a simple model consisting of a bulk 
scalar with higher derivative ``interaction'' terms
\begin{eqnarray}
\label{eq:saction}
\nonumber
S &=& {1\over 2}\int d^5 X \sqrt{G} (\partial\Phi)^2 + {\lambda_2\over
2} \int d^5 X \sqrt{G}\left(\Box \Phi\right)^2\\
& & {} + \sum_{n=3}^\infty{\lambda_n\over 2} \int d^5 X\sqrt{G} \Phi
\Box^{n} \Phi,
\end{eqnarray}
where $\Box$ is the scalar Laplacian on AdS.  
%In five dimensions, $[\Phi]=3/2$ and $[\lambda_n] =2-2n$.  
First we examine an ${\cal O}(\lambda_2)$ correction to the 
two-point function of $\Phi$.  
\begin{equation}
\langle \Phi(X)\Phi(X')\rangle \sim  D(X,X') + \lambda_2
{\delta^5(X-X')\over\sqrt{G}}.
\end{equation}
In this equation, $D(X,X')$ is the scalar propagator derived from the free scalar action.  Performing a Fourier transform along the 4D coordinates $x^\mu$, this becomes (with $p^\mu$ the coordinate momentum, $p\cdot x=\eta_{\mu\nu} p^\mu
x^\nu$, and $p=\sqrt{\eta_{\mu\nu} p^\mu p^\nu}$)
\begin{eqnarray}
\nonumber
G_p(z,z') &\equiv& \int d^4 x e^{i p\cdot x} \langle \Phi(x,z)\Phi(0,z')\rangle\\
&\sim& D_p(z,z') + \lambda_2 (kz)^5\delta(z-z'), 
\end{eqnarray}
Projecting onto the $m\mbox{-th}$ KK
wavefunction,  $\psi_m(z)$, we find 
\begin{equation}
\int_{1\over k}^{1\over T} {dz'\over (kz')^3}\psi_m(z') G_p(z,z')\sim\psi_m(z)\left[{1\over p^2+ m_n^2} + {\lambda_2} (kz)^2\right],
\end{equation}
where we have used the representation of the propagator in terms of
modes, as well as the orthogonality condition for scalar KK
wavefunctions ($m_n$ is the mass of the $n\mbox{-th}$ KK state).
It is natural to take $\lambda_2$ to scale as $\lambda_2\sim
M_5^{-2}$, where $M_5$ is the fundamental scale in 5D.  For fixed $n$
and $p\gg m_n$, the $\lambda_2$ correction to the propagator becomes
leading when 
\begin{equation}
\label{eq:slide}
p \sim {M_5/kz}.
\end{equation}
This signals the breakdown of the EFT description
for the given mode. Eq.~(\ref{eq:slide}) is simply the well known
result~\cite{RS1} that the local cutoff scales with $z$. 
Here we have made clear the relevance of the sliding cutoff for particular modes, and the importance of the localization of the mode to the breakdown of the EFT becomes apparent.  From this estimate, it is clear that the loss of predictivity has nothing to do with loop effects, or with the nature of a specific choice of regulator. 

An  estimate based on a single insertion of the $\lambda_2$ coupling
is insufficient for the correlator involving only external KK zero modes.
In this case insertions higher derivatives operators are necessary.  Consider the 1PI zero mode two-point correlator in the presence of all the higher dimension
terms in Eq.~(\ref{eq:saction}) 
 \begin{equation}
G^{-1}(p) = p^2 + \lambda_2 p^4 
+\sum_{n=3} {\hat \lambda}_n \left({k\over M}\right)^{2n-2} 
{T^4\over k^2} \left({p\over T}\right)^{2n},
\end{equation}
with ${\hat\lambda}_n\sim M^{2n-2}\lambda_n$.  In order to have a sensible low energy theory, the $p^2$ term must dominate the corrections from operators with more derivatives.  This means that we must have 
\begin{equation}
{p\over T} < {M\over k} \left({k\over T}\right)^{1/(n-1)},
\end{equation}
where constants of order unity have been dropped.  
Interestingly, operators with more derivatives 
dominate at lower energy scales, and an operator 
with $n\gg 1$ becomes as important as the 
zeroth order propagator at a scale $p\sim T$.

From this analysis, we conclude that we should not expect to be
able to calculate zero mode observables (and in particular RG flows) for momenta much larger than a TeV, contrary to what was proposed in~\cite{rsrun}.
Instead, if we are interested in calculating at such large momenta we
must restrict ourselves to observables localized to the Planck brane.

\section{Gauge fields in AdS}

We will study scalar QED propagating in the curved background of
Eq.~(\ref{eq:metric})
\begin{equation}
S=\int d^5 X\sqrt{G} \left[{1\over 4 g_5^2} F_{MN} F^{MN} + |D_M\Phi|^2\right].
\end{equation}
First, consider the gauge field zero mode propagator.  Performing the
computation explicitly in 5D, we find using dimensional regularization\footnote{As we will show in a subsequent paper~\cite{us},
the finite non-analytic parts of this quantity can be computed in a
regulator independent manner.} and working in $A_5=0$ gauge (irrelevant constants have been dropped)  
\begin{eqnarray}
\label{eq:adspi}
\nonumber
\Pi_{\mbox{\tiny{AdS}}}(q^2) &=& {1\over g_4^2}+{1\over
48\pi^2}\left[{1\over\epsilon}- \ln\left({q^2\over kT}\right) +
{1\over 2}\ln\left({\mu^2\over kT}\right)\right]\\
& & {}  - {1\over 16\pi^2} \int_0^1 dx x\sqrt{1-x^2} \ln N\left({\sqrt{q^2} x\over 2}\right)
\end{eqnarray}
where $N(p) = I_1(p/T) K_1(p/k) - I_1(p/k) K_1(p/T)$, with
$I_1(x),K_1(x)$ modified Bessel functions of order one, and as in flat
space $1/g_4^2 = R/g_5^2$, 
with $R$ the proper distance between the boundaries.  As on the
flat orbifold, it can be shown that the $1/\epsilon$ pole represents
logarithmic divergences that renormalize gauge kinetic terms localized
on the boundary branes at $z=1/k$ and $z=1/T$~\cite{us}.  The coefficient of the pole even matches that of Eq.~(\ref{eq:orb}), leading to the same RG
equations for the boundary terms.  Physically, this is to be expected,
since the UV divergences of field theory in curved space arise from
distances shorter than the curvature scale and are therefore identical
to those in flat space.  In particular, the RG flows for couplings to
operators that are present in both flat and curved space should be the
same\footnote{However, it is possible for new operators composed from
powers of the background curvature to play a role in the RG flows.}.
It follows that as in flat space, the 5D bulk gauge coupling does not
run.

For $\sqrt{q^2}\ll T,$ the zero mode of the scalar dominates
Eq.~(\ref{eq:adspi}) giving rise to low energy logarithmic energy
dependence.  For $\sqrt{q^2}\gg T$ the behavior is power law 
\begin{equation}
\Pi_{\mbox{\tiny{AdS}}}(q^2)\sim {\sqrt{q^2}\over T}
\end{equation}
and the zero mode observable becomes strongly coupled, in accord with
the analysis of the previous section.  In fact, one does not have to consider loop effects to see power-law behavior.  The non-renormalizable operator
\begin{equation}
\label{eq:tree}
S=\cdots + {\lambda\over 2}\int d^5 X\sqrt{G} F_{MN}\Box F^{MN}
\end{equation}
will give rise to an analytic power-law contribution to Eq.~(\ref{eq:adspi}) of the form $\Pi_{\mbox{\tiny{AdS}}}(q^2)\sim\cdots + k \lambda q^2/T^2$.  In~\cite{rsrun}, it was argued that the zero mode gauge propagator can be studied at energies much higher than the KK scale, provided a suitable momentum cutoff is employed.  Because the contribution from Eq.~(\ref{eq:tree}) is a tree-level effect, it is clear the breakdown of the zero mode observable cannot be avoided by a mere choice of regulator.

Although we cannot use the zero mode observables to study high energy
gauge theory, the Planck brane correlator is still perturbative at
energies much larger than the KK scale.  To see that loop corrections
to this quantity are logarithmic for a large range of scales, consider
the gauge propagator (in $A_5=0$ gauge) with one point on the Planck
brane for external momenta larger than $T$:
\begin{equation}
\label{eq:exp}
D_{p;\mu\nu}(z,1/k)\simeq {k z\over p}{K_1(pz)\over K_0(p/k)}\eta_{\mu\nu}+\mbox{pure gauge}.
\end{equation}
For $z$ near $1/T$, $pz\gg 1$, and $K_1(pz)\sim\sqrt{\pi/(2pz)} \exp{(-pz)}$.  We then see that this quantity has almost no overlap with the excited KK states of the bulk scalar, which are localized towards the TeV brane.  It follows that the contribution from these states to the one-loop Planck correlator, which could potentially give rise to a power law of the form $\sqrt{q^2}/T$ is
practically zero (it is supressed by powers of $T/k\ll 1$).  However, because the scalar zero mode profile is flat, it does not suffer from the exponential suppression, and gives the dominant contribution to the correlator.  We thus expect the zero mode logarithm to be the leading energy dependence of the one-loop Green's function.  As the external momentum reaches the curvature scale $k$, the heavy KK modes which have support near the Planck brane start to contribute.  These modes give rise to power law behavior that becomes dominant near the scale $k$.  

Instead of performing the 5D calculation just described, we will show
how to use AdS/CFT duality as it applies to the RS model to obtain
quantitative information about the one-loop Planck correlator.  Our toy gauge theory in AdS corresponds to the 4D theory of a $U(1)$ gauge field, a  charged scalar, and a CFT explicitly broken by a UV cutoff (the dual of the Planck brane) and spontaneously broken in the IR (the TeV brane in the 4D context)\cite{APR,RZ}.  The gauge field couples minimally to the scalar, and couples weakly to an anomaly free $U(1)$ subgroup of the global symmetries of the CFT.  By
similar arguments as those of~\cite{Gubser} for the graviton, the presence
of the Planck brane in the 5D theory renders the KK zero mode of the bulk
scalar normalizable, and therefore implies the existence of a 4D
massless scalar which couples to the CFT through a dimension four
operator.   (This can be verified by checking that the KK corrections
to scalar exchange on the Planck brane in the 5D theory match the contributions to the force law from the CFT in the 4D dual).  Thus our 5D theory has a 4D dual description (ignoring couplings to 4D gravity) 
given by
\begin{eqnarray}
\nonumber
{\cal L}_{\mbox{\tiny{4D}}} &=& {\cal L}_{\mbox{\tiny{CFT}}}  +
{1\over 4 g^2} F_{\mu\nu} F^{\mu\nu} 
+  A_\mu J_{\mbox{\tiny{CFT}}}^\mu \\
& & {} + |D_\mu\phi|^2 + c\left(\phi{\cal O}_4 +\mbox{h.c.}\right),
\end{eqnarray}
where $c$ is a coupling of order $1/k$.  While we do not know precisely what CFT is dual to the RS model, it is still possible to use this Lagrangian to compute Planck brane correlators of bulk fields.  For instance, the Planck brane vacuum polarization is given by 
\begin{eqnarray}
\label{eq:cftpi}
\Pi^{\mu\nu}(q^2) &=& \left(q^\mu q^\nu - q^2\eta^{\mu\nu}\right)\left[{1\over g^2}  - {1\over 48\pi^2} \ln\left({q^2\over\mu^2}\right)\right]\\
\nonumber
& & {} + \int d^4 x e^{i q\cdot x} \left\langle J^\mu_{\mbox{\tiny{CFT}}}(x) J^\nu_{\mbox{\tiny{CFT}}}(0)\right\rangle_{\mbox{\tiny{CFT}}} +{\cal O}(|c|^2). 
\end{eqnarray}
The first line is simply the contribution of a 4D scalar to the vacuum
polarization.  The second line is the correction to the correlator due
to pure CFT effects.  Writing   
\begin{equation}
\label{eq:JJ}
\int d^4 x e^{i q\cdot x} \left\langle J^\mu_{\mbox{\tiny{CFT}}}(x) J^\nu_{\mbox{\tiny{CFT}}}(0)\right\rangle_{\mbox{\tiny{CFT}}}=\left(q^\mu q^\nu - q^2\eta^{\mu\nu}\right)\Pi(q^2),
\end{equation} 
it can be shown~\cite{APR} that for $\sqrt{q^2}\gg T$ (note that the effects of
being in a vacuum state with broken conformal symmetry are suppressed
exponentially at energies much larger than $T$)
\begin{equation}
\Pi(q^2)\simeq{1\over 2 g_5^2 k}\ln\left({q^2\over k^2}\right).
\end{equation}
In the AdS description, this term corresponds to the contribution of
the KK modes of the gauge field to the tree level Planck correlator.
Because in the bulk theory Eq.~(\ref{eq:JJ}) represents a tree level
effect, it gives the leading behavior of the correlator.  In settings where the AdS/CFT correspondence has been well tested, the
CFT is a large $N$ $SU(N)$ gauge theory, so Eq.~(\ref{eq:JJ}) is
expected to be larger than the scalar correction by powers of 
$N$.   However, the coefficient of this
classical logarithm is universal, so pure CFT effects
will cancel in predictions for the difference of low-energy couplings
in terms of high energy parameters. 

Besides the terms explicitly shown in Eq.~(\ref{eq:cftpi}), there are also corrections which involve insertions of the operator ${\cal O}_4$.  These terms are suppressed by powers of $k$, and correspond to the exponentially damped scalar KK corrections to the 5D correlator.  They can be calculated in terms of CFT correlators (in the vacuum with broken conformal symmetry) of ${\cal O}_4$ and $J^\mu_{\mbox{\tiny{CFT}}}$.  Such correlators can be obtained~\cite{us} using the rules of~\cite{AdSCFT} for relating CFT correlators to solutions of classical AdS field equations.    

So far we have only considered the evolution of the Planck brane
vacuum polarization due to one-loop effects of bulk matter.  It is
easy to see that brane localized fields will also contribute
logarithms with the usual 4D beta function coefficients.  The effects
of Planck localized fields persist for all energies up to the
curvature scale.  However, for $\sqrt{q^2}\gg T$, the contribution due to TeV
brane fields on the Planck brane correlator is suppressed by an
exponential of $\sqrt{q^2}/T$, since for high energies this is a highly
non-local effect (explicitly, the suppression is by two powers of
Eq.~(\ref{eq:exp}) evaluated at $z=1/T$).  This can be understood also
in the 4D dual description~\cite{APR,RZ}.  There, Planck brane fields are
spectators to the CFT dynamics, so it is clear that they give rise to
the usual 4D vacuum polarization effects.  On the other hand, TeV
brane localized fields arise in the 4D dual as condensates of CFT
states in the IR.  These bound states do not contribute to the 
running of the couplings above TeV scale energies.

Although we only explicitly considered a toy scalar model, we expect
that loops of bulk non-abelian gauge and fermion fields will also
generate logarithmic corrections to the Planck correlator with the
usual 4D coefficient.  Because  at energies below the KK scale the
Planck correlators match on to the Green's functions of zero
mode fields, gauge coupling evolution in the RS model with Standard
Model gauge fields in the bulk is generically similar to the situation
in the minimal (4D) Standard Model with an energy desert.  It is
therefore possible to make perturbative predictions for the low energy
couplings of the model in terms of the parameters in the underlying 5D
Lagrangian.  

\section{Conclusions}

It is clear that effective field theory in curved backgrounds is more subtle than in flat space. We have seen that in truncated AdS geometries, we can determine the scale at which effective field theory breaks down for classes of correlators.  For instance, it is not possible to calculate above the TeV scale for
zero mode correlators, but correlators whose end points are
restricted
to the boundary (``Planck'') brane do succumb to field theoretic
techniques reliably, all the way up to the curvature scale. Furthermore, there are certain observables, e.g. differences of gauge couplings, which can be calculated without detailed knowledge of the dual CFT.  As was previously postulated~\cite{rsrun}~\cite{pomarol,choi}, these couplings run logarithmically.

This leads to the intriguing possibility, first noted by~\cite{rsrun}, that a version the RS proposal for addressing the hierarchy problem with bulk Standard Model gauge fields could also predict coupling constant unification within a
grand unified context\footnote{In the proposal of~\cite{pomarol}, the
Standard Model is localized to the Planck brane, so supersymmetry must
be employed to stabilize the weak scale.  In that case, the AdS warp
factor is used to generate an exponentially low supersymmetry breaking
scale.}.  In order to maintain the hierarchy in the presence of bulk
gauge fields, the Higgs scalar responsible for electroweak symmetry breaking must be confined to the TeV brane.  It only contributes to gauge coupling evolution up to the TeV scale.  Standard Model fermions may propagate either in the bulk or on the TeV brane.  However, since the fermions come in
complete $SU(5)$ multiplets, the RS prediction for low energy coupling
constant relations is model independent, and qualitatively similar to
that of the Standard Model.  Achieving unification, though, could necessitate additional, perhaps ad hoc, physics since the running above the TeV scale excludes the Higgs.  As in the Standard Model, there are potential threshold corrections to logarithmic evolution near the  GUT scale.  There are also TeV threshold corrections that arise from the matching of the Planck to the zero mode Green's functions.  Although these corrections are not accessible to our 4D CFT
calculation, they can be unambiguously calculated in 5D.
The details for plausible models have yet to be worked out.

I.R. would like to acknowledge the LBL theory group for its hospitality.  The work of I.R. was supported in part by the DOE contracts DOE-ER-40682-143 and DEAC02-6CH03000.  W.G. was supported in part by the DOE contract DE-AC03-76SF00098 and by the NSF grant PHY-0098840.

\end{multicols}
\end{document}